# Electronic structure and transport properties of $La_{0.7}Ce_{0.3}MnO_3$

W. J. Chang[1], J. Y. Tsai[2], H.-T. Jeng[3], J.-Y. Lin[2], Kenneth Y.-J. Zhang[4], H. L. Liu[4], J. M. Lee[5], J. M. Chen[5], K. H. Wu[1], T. M. Uen[1], Y. S. Gou[1], and J. Y. Juang[1]

[1]Department of Electrophysics, National Chiao Tung University, Hsinchu 30050, Taiwan
[2]Institute of Physics, National Chiao Tung University, Hsinchu 30050, Taiwan
[3]Physics Division, National Center for Theoretical Sciences, Hsinchu 30013, Taiwan
[4]Department of Physics, National Taiwan Normal University, Taipei 11677, Taiwan
[5]National Synchrotron Radiation Research Center (NSRRC), Hsinchu 30076, Taiwan

X-ray absorption spectroscopy (XAS), optical reflectance spectroscopy, and the Hall effect measurements were used to investigate the electronic structure in $La_{0.7}Ce_{0.3}MnO_3$ thin films (LCeMO). The XAS results are consistent with those obtained from LDA+$U$ calculations. In that the doping of Ce has shifted up the Fermi level and resulted in marked shrinkage of hole pockets originally existing in $La_{0.7}Ca_{0.3}MnO_3$ (LCaMO). The Hall measurements indicate that in LCeMO the carriers are still displaying the characteristics of holes as LDA+$U$ calculations predict. Analyses of the optical reflectance spectra evidently disapprove the scenario that the present LCeMO might have been dominated by the La-deficient phases.

Over the last decade, researches on the rare-earth manganese perovskites, $RE_{1-x}A_xMnO_3$ (RE: rare-earth ion, A: alkaline earth cation), have been largely devoted to the characteristics related to the electric transport and magnetic properties, e.g., colossal magnetoresistance, phase separation, and the competition between the charge, spin, and orbital order parameters.[1] In these compounds, the charge state of Mn displays a mixed-valence characteristic of $Mn^{3+}$ and $Mn^{4+}$, and the ratio of $Mn^{3+}/Mn^{4+}$ depends strongly on the doping level of the divalent cation. Intuitively, substituting the divalent cation with the tetravalent cation, such as Ce, Sn, etc., should impel the valence state of Mn to a mixed-valence state of $Mn^{2+}$ and $Mn^{3+}$.[2,3] It then raises curiosities on how the electric transport and magnetic properties of tetravalent cation doped manganites would prevail, as compared with the divalent-doped system.

Previously, the transport properties and the electronic structure of LCeMO had been reported with results obtained from the tunneling junction, photoemission, and x-ray absorption spectroscopy (XAS) experiments.[4-8] The results showed that the valence state of Mn is indeed a mixed-valence of $Mn^{2+}/Mn^{3+}$, and it is probably electron-doped. However, there are discrepancies among the reported results. For instance, the tunneling junction experiments suggested that the itinerant carriers in LCeMO are the minority spin carriers[4], which was in sharp contrast to the conclusion of the majority spin carriers drawn from XAS experiments and theoretical calculations.[8,9] On the other hand, some reports even proposed that it is the La-deficient phase existent in LCeMO that gives rise to the metal-insulator transition.[10] In this paper, we have utilized various independent experimental and theoretical approaches to systematically examine the LCeMO system. Our results unambiguously clarified some of the outstanding controversies in this system.

Details of preparing the single phase LCeMO on (100) $SrTiO_3$ substrates by pulsed-laser deposition and the associated structure-property analyses were reported in Ref. 11. The x-ray scattering and x-ray diffraction results indicated that the obtained LCeMO films were highly epitaxial single-phase samples with negligible traces of impurity phases like $CeO_2$ and MnO. The O $K$-edge and Mn $L$-edge XAS spectra were carried out using linear polarized synchrotron radiation from a 6-m high-energy spherical grating monochromator beamline located at NSRRC in Taiwan. Details of XAS experiments can be found in Ref. 12. Being different from previous XAS measurements at 300 K[6-8], the O 1s XAS spectra of LCeMO and LCaMO were taken by x-ray fluorescence yield at 30 K, which directly probed the electronic structure in the ferromagnetic state for both samples. The longitudinal resistivity $\rho_{xx}$ and Hall resistivity $\rho_{xy}$ were measured by the four-probe and six-probe techniques, respectively. All of $\rho_{xx}$, $\rho_{xy}$, and the magnetization $M$ were measured in an 8-Tesla QUANTUM DESIGN® physical properties measuring system (PPMS). Near-normal optical reflectance spectra were measured over a broad frequency range (from 50 to 52000 $cm^{-1}$) at temperatures between 10 and 330 K. The thin-film optics and the Drude-Lorentz analysis were used to modeling the optical properties of these samples.[13] The band structure calculations were performed using the full-potential projected augmented wave method[14] as implemented in the Vienna $ab$ $initio$ simulation package (VASP)[15] within the local-density approximation plus on-site Coulomb interaction $U$ (LDA+$U$) scheme.[16] In the LDA+$U$ calculations, we used Coulomb energy $U$=5.0 eV and exchange parameter $J$=0.95 eV for Ce(La)-4$f$ electrons, while $U$=4.0 eV and $J$=0.87 eV were used for Mn-3$d$ electrons.[8]

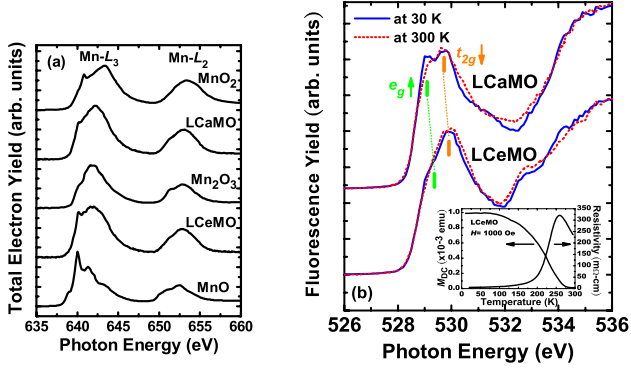

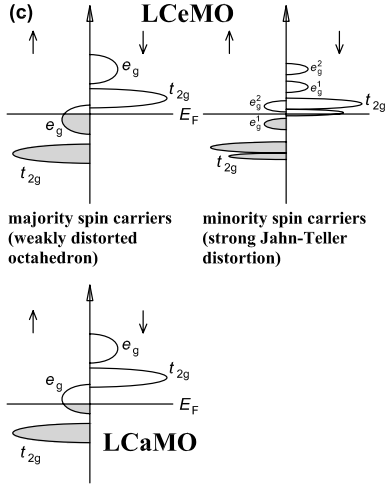

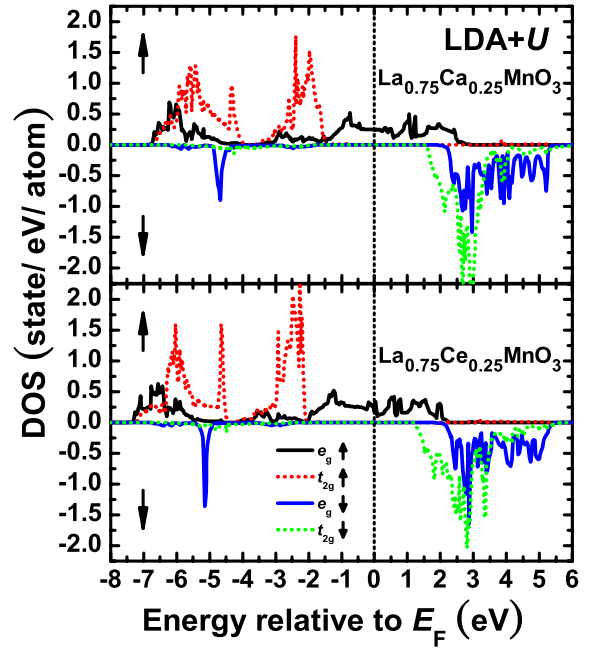

FIG. 1: (a) Spectra of Mn $L$-edge XAS for Mn-ions with various valence states by total electron yield. (b) Spectra of O $K$-edge XAS (by florescence yield) of LCaMO and LCeMO at 300 and 30 K, respectively. The absorption peaks at 529 and 530 eV were assigned to the unoccupied states of $e_g\uparrow$ and $t_{2g}\downarrow$ orbitals, respectively. Inset: $\rho(T)$ and $M(T)$ of LCeMO, showing the typical CMR magnetotransport characteristics. (c) Left: Schematic band diagrams describing the majority spin scenario for LCeMO (upper panel) and LCaMO (lower panel). Right: Schematic diagram, quoted from Ref. 4, depicting the minority spin carriers scenario in LCeMO.

To characterize the valence state of manganese in LCeMO, Mn $L$-edge XAS spectra of $MnO_2$, $Mn_2O_3$, MnO, LCaMO, and LCeMO were measured and are shown in Fig. 1(a) for comparison. The spectral weight of Mn $L$-edge for LCeMO evidently exhibits the characteristics of both $Mn^{2+}$ and $Mn^{3+}$, which has been previously interpreted as a manifestation of $Mn^{2+}/Mn^{3+}$ mixed-state in LCeMO.[6-7] In Fig. 1(b), O $K$-edge XAS spectrum of LCeMO reveals a shoulder near absorption edge as compared with the first pronounced peak of LCaMO at 529 eV. These results suggest that there are fewer unoccupied states in $e_g\uparrow$ band of LCeMO than in that of LCaMO due to extra electrons doped into the $e_g$ band of LCeMO, presumably originated from substituting $Ca^{2+}$ with $Ce^{4+}$. Based on these XAS results, the electronic structure of LCeMO appears to be consistent with the scenario of the majority spin carriers associated with strong Hund's rule effects. On the other hand, for the minority spin scenario to prevail one would expect otherwise a strong O $K$-edge spectral weight in the vicinity of the absorption edge. The electronic structure diagrams for the scenarios discussed above are depicted schematically in Fig. 1 (c). The density of states spectra obtained from LDA+$U$ calculations (Fig. 2) also revealed that both LCeMO and LCaMO are majority spin half-metals with fewer unoccupied $e_g\uparrow$ states in LCeMO. This is in good agreement with the observed XAS spectra mentioned above. The above analyses unambiguously demonstrate that electrons are indeed doped into LCeMO. Moreover, as displayed in the inset of Fig. 1(b), the mixed-valence $Mn^{2+}/Mn^{3+}$ also leads to typical magneto-transport properties of CMR manganites, with $T_c$ and $T_{IM}$ being around 260 K, which is also consistent with most of the previous reports.[2-8]

FIG. 2: DOS of Mn $3d$ in LCaMO and LCeMO calculated by LDA+$U$ method. The parameters used in these calculations are described in the text.

The next question of whether the doped electrons actually drive the itinerant carriers to become *electrons* remains to be confirmed. To examine this scenario, the Hall measurements are indispensable. In ferromagnets, the Hall resistivity can be expressed by $\rho_{xy}(B, T) = R_H(T) B + \mu_0 R_S(T) M(B, T)$, where $R_H$ is the ordinary Hall coefficient, $R_S$ the anomalous Hall coefficient, $\mu_0$ the vacuum permeability, and magnetic induction $B=\mu_0 [H+(1-N_d) M]$ with the demagnetization factor $N_d\sim1$ in our film geometry. In order to improve the accuracy of the experimental data, values of $\rho_{xy}$ were obtained by averaging the results acquired from two scans of $B$. The field scans range from 8 T to -8T. Combining with the measurements of $M(B)$, $R_H$ and $R_S$ were then obtained by fitting $\rho_{xy}(B)$ to the above-mentioned Hall resistivity equation. As shown in Fig. 3, the positive slope $d\rho_{xy}/dB$ in the high field regime, where $M(B)$ saturates, strongly suggests that the itinerant carriers in the present LCeMO films, like that in LCaMO films, are *holes*. In addition to the ordinary Hall effect discussed above, the extraordinary Hall effect is generally associated with large and localized magnetic moments existent in the material.[17] As is evident in Fig. 3, LCeMO appears to have a more significant extraordinary Hall effect that LCaMO even at 10 K. The difference might be due to strain-induced effects and could be either intrinsic (ion size difference) or extrinsic (extent of lattice mismatch with substrate). It might also be due to the difference in electronic bandwidth for the itinerant carriers, which in turn will affect the efficiency of ferromagnetic transition via the double-exchange mechanism.

We now return to the issue of the p-type behavior revealed by Hall measurements. Starting from the ionic band structure of parent LaMnO$_3$, due to the strong Jahn-Teller effect, the $e_g\uparrow$ band splits into two sub-bands, $e_g^1\uparrow$ and $e_g^2\uparrow$ and results in the commonly observed insulator behavior.[18, 19] Therefore, electron doping in LCeMO presumably could have led to electron carriers in this model. To unravel the origin of why the carriers are still holes in LCeMO, we performed band structure calculations for both the hole-doped LCaMO and electron-doped LCeMO by using the LDA+$U$ method. As shown in Fig. 4, LCaMO is clearly a p-type metal with significant hole pockets near the Fermi surface as expected. Surprisingly, the band dispersions of LCeMO evidently also show nontrivial hole pockets around the Fermi surface, in spite of the electron doping from Ce$^{4+}$ ions. Nonetheless, the Fermi level has been raised significantly in the latter case. It has been pointed out that the strong Jahn-Teller distortion due to the local lattice distortion of Mn$^{3+}$ in LaMnO$_3$ is largely exempted in both of the Mn$^{3+}$/Mn$^{4+}$ and Mn$^{2+}$/Mn$^{3+}$ cases.[1] Consequently, the splitting between $e_g^1$ and $e_g^2$ induced by the Jahn-Teller effect is gradually diminished when LaMnO$_3$ is doped with either divalent Ca or tetravalent Ce. The evolution of the electronic structure of La$_{1-x}$Sr$_x$MnO$_3$, which relates Jahn-Teller distortion to the hole doping level, has demonstrated the closing of the band gap between $e_g^1\uparrow$ and $e_g^2\uparrow$ sub-bands with increasing doping.[19] We believe that it is the similar effects that lead the Fermi level of both LCaMO and LCeMO to lie in the $e_g\uparrow$ band, making the transport carriers exhibit hole-like behaviors in both cases.

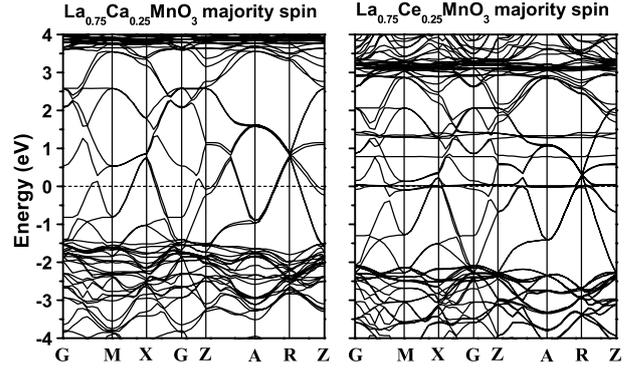

**FIG. 4:** The results of LDA+$U$ band structure calculations for LCaMO and LCeMO. Only the majority spin part is shown.

The next issue to be addressed is whether or not all of these are just the manifestations of La-deficient phase. Since it is rather difficult to make clear-cut distinction between them by merely relying on magneto-transport and XAS measurements, we performed independent optical reflectance measurements to provide additional information for this purpose. Figure 5 shows the measured optical reflectance of the (100) SrTiO$_3$ substrate at 300 K and the LCeMO, La$_{0.7}$MnO$_3$ film (as grown) at 20 K. Both LCeMO and La$_{0.7}$MnO$_3$ film have approximately the same Curie temperature around 260 K. here are several important features to these spectra. First, the far-infrared spectrum of the LCeMO film can be described by a weak, overdamped Drude contribution typical of a poor conductor. As a consequence, the effect of multiple reflection of light in the film on the substrate is clearly seen in the mid-infrared frequency region. Second, the spectral weight of the infrared reflectance of the La$_{0.7}$MnO$_3$ film is substantially increased, indicative of increased carrier concentrations. Third, from the Hall measurements and optical data, we derive the carrier effective mass of the three manganites studied by the relation $\omega_p^2 = \frac{4\pi \cdot e}{m^* \cdot R_H}$. As listed in Table I, the ratio of the effective mass $m^*/m^*_{(LCaMO)}$ are 1.62 and 0.398 for LCeMO and La$_{0.7}$MnO$_3$ film, respectively. Consequently, LCeMO is apparently having much heavier holes and appears to be a narrow band manganite as compared to LCaMO. This is in contrast to that implied by the result of $m^*/m^*_{(LCaMO)}$ for La$_{0.7}$MnO$_3$ film shown in Table I, and virtually rules out the scenario that the LCeMO films studied in the present work is of La-deficient nature. Although Yanagida et al.[10] reported the co-existence of nano-cluster cerium oxides with minor phases of La-deficient La$_{1-x}$MnO$_3$ and A-site cation-deficient (La, Ce$_{1-\delta}$)MnO$_3$ in their post-annealed films which showed similar metal-insulator transition, their as-grown film, albeit in single phase, was an insulator. This is different from the present as-grown LCeMO films, which are metallic even without post-annealing. The differences are presumably due to different film growth processes.

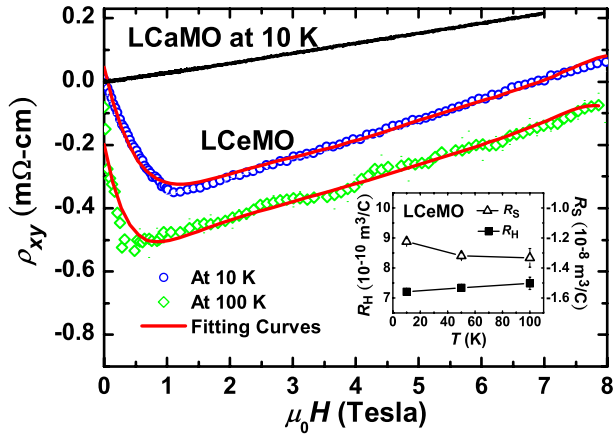

**FIG. 3:** The Hall measurements of LCeMO and LCaMO. The lines are from the fits to the equation mentioned in the text. Inset: the $T$ dependence of $R_H$ and $R_S$.

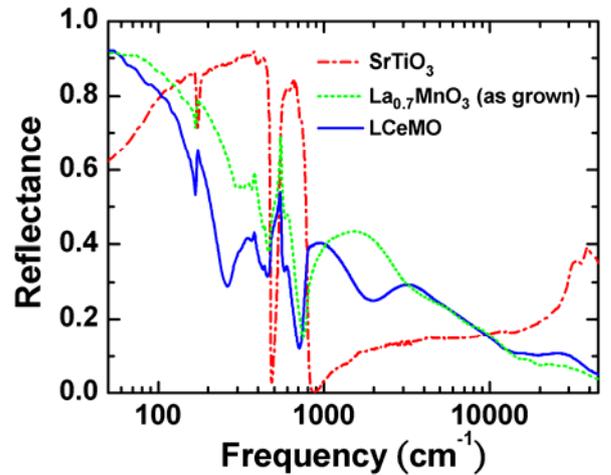

**FIG. 5:** The reflectance spectra of the LCeMO and an as-grown La$_{0.7}$MnO$_3$ film measured at 20 K. For comparison, the room-temperature reflectance data of the (100) SrTiO$_3$ substrate is also included.

|  | LCeMO | LCaMO | La$_{0.7}$MnO$_3$ |
|---|---|---|---|
| $R_H$ (m$^3$/C) (at 10 K) | 7.21×10$^{-10}$ | 3.16×10$^{-10}$ | 8.64×10$^{-10}$ |
| $\rho$ (mΩ-cm) (at 10 K) | 5.09 | 0.366 | 1.28 |
| $\omega_p$ (1/s) (at 20 K) | 4.71×10$^{14}$ | 9.08×10$^{14}$ | 8.63×10$^{14}$ |
| $m^*/m^*_{(LCaMO)}$ | 1.62 | 1 | 0.398 |

**TABLE I:** Summary of $R_H$, $\rho$, $\omega_p$, and $m^*/m^*_{(LCaMO)}$ for listed samples.

In summary, we have presented the detailed results of Mn $L$-edge and O $K$-edge XAS which suggest that at low temperatures LCeMO is a majority spin carrier ferromagnet. The results also display clear evidence of electron-doping into the $e_g \uparrow$ sub-band of LCeMO. However, both Hall measurements and LDA+$U$ band structure calculations indicate that the doped electrons did not drive LCeMO into a n-type manganite and the itinerant carriers are still holes, but with a much less concentration as compared with that of LCaMO. Finally, the results of optical reflectance spectra, together with the Hall measurements, evidently dismiss the scenario that the present as-grown LCeMO films are dominated by La-deficient phases.

This work was supported by the National Science Council of Taiwan, under grants: NSC 93-2112-M-009-015, 93-2112-M-009-016, 94-2119-M-007-001, and 93-2112-M-003-005.